\documentclass[twocolumn,aps,pre]{revtex4}
\usepackage[dvips]{graphicx}
\usepackage{array}
\usepackage{amsmath}
\usepackage{amssymb}

\usepackage{color}

\begin{document}

\title{Impact of spherical nanoparticles  on the nematic order parameters}
\author{C. Kyrou$^1$, S. Kralj$^2$, M. Panagopoulou$^3$, Y. Raptis$^3$, G. Nounesis$^4$, I. Lelidis$^{1}$}
\affiliation{$^1$ Faculty of Physics, National and Kapodistrian University of Athens, \\
Panepistimiopolis, Zografos, Athens 157 84, Greece\\
$^2$ Faculty of Natural Sciences and Mathematics, University of Maribor, 2000 Maribor, Slovenia\\
$^3$ Physics Department, National Technical University of Athens, Heroon Polytechniou 9, 15780 Zographou,
Athens, Greece\\
$^4$ Biomolecular Physics Laboratory, National Centre for Scientific
Research “Demokritos,” Aghia Paraskevi, Greece
}
\date{\today}

\begin{abstract}
We study experimentally the impact of spherical nanoparticles on the orientational order parameters of a host nematic liquid crystal. We use spherical core-shell quantum dots that are surface functionalized to promote homeotropic anchoring on their interface with the liquid crystal host. We show experimentally that the orientational order may be strongly affected by the presence of spherical nanoparticles even at low concentrations.
The orientational order of the composite system is probed by means of polarised micro-Raman spectroscopy and by optical birefringence measurements as function of temperature and concentration. Our data show that the orientational order depends on the concentration in a non linear way, and the existence of a crossover concentration $\chi_c \approx 0.004\, pw$. It separates two different regimes exhibiting pure-liquid crystal like ($\chi <\chi_c$) and distorted-nematic ordering ($\chi >\chi_c$), respectively. In the latter phase the degree of ordering is lower with respect to the pure-liquid crystal nematic phase.
\end{abstract}

 \maketitle

\section{Introduction}

 Liquid crystals (LC) combine the fluidity of ordinary liquids with (quasi-) long range order and exhibit anisotropic properties on a macroscopic range \cite{degennes,chandra,lev,onsager} that give them, among others, rapid response in external fields. Nematic liquid crystal displays are based exactly on these properties \cite{display,display2}, and  on the crucial role of interfaces as well \cite{sandro, paolo}. Mixtures of LC with colloidal particles have attracted the attention of researchers long time ago \cite{stark,abbott2}. During the last two decades, hybrid systems composed by liquid crystals doped with nanoparticles (NP) have been widely investigated for their new and/or enhanced properties such as electro-optical \cite{kaur,modulator}, phase transitions \cite{SmOrder}, stabilization  and phase separation \cite{yoshida,eva,TGB}, topological defects \cite{Manu,abbott}, instabilities, photonic LC \cite{photonic}, anchoring \cite{anchoring}, etc. For this reason, the interaction of the NP with the host matrix is of particular interest.
 Nematic liquid crystals (NLC) are characterised by orientational order along a common direction called nematic director $\mathbf{n}$ (with $\mathbf{n}^2=1$ and $\mathbf{n}\equiv -\mathbf{n}$). The mesogenic molecules align around the director in the mean, and the quality of their alignment is quantified by the orientational order parameter. Since most of NLC applications are based on their anisotropic properties, such as dielectric anisotropy and  birefringence, their quality depends on the degree of alignment around the director.

 The most common techniques that permit the calculation of the orientational order parameter are dielectric permittivity, birefringence, absorption, infrared spectroscopy, polarised fluorescence, Electron Paramagnetic Resonance, X-ray and neutron diffraction, Electronic and Vibrational Spectrocopy, Polarized Raman Scattering (PRS) etc.
 Raman spectroscopy is a powerful tool for the study of soft matter systems that  gives access to the orientational order of liquid crystals. In particular, Raman peaks provide information on molecular vibrations and their local environment. Therefore, one can deduce information concerning the packing and local order in the liquid crystal phase. For hybrid systems, the interaction between colloidal or nano-particles and the liquid crystal matrix, can be investigated by PRS. Although Raman spectroscopy is a technically complex method, it has the advantage to give access to the first two moments,  $\left\langle P_2\right\rangle$ and  $\left\langle P_4\right\rangle$, of the orientational distribution function \cite{luckhurst,Jen,deJeu,Miyano,spiridonov}. Therefore it has been used for the last four decades since the seminal paper  of Jen, Clark, Pershan and Priestley (JCPP) \cite{Jen} that laid to the foundation of the PRS technique in LCs. New variants of the original method adapted for complex matter have been developed recently \cite{SERS,Kato,Helen,confocal,smalyukh,phasemod,SERSgraph,ferro,tb}.
LCs are used as solvents to control the order of anisotropic in shape particles, such as carbon nanotubes \cite{CNTubes,samo,eric}, and nanoplates \cite{graphene,graphBP}, as well as particles with special properties, e.g., ferromagnetic \cite{creanga}, and ferroelectric particles \cite{li,Reznikov,freederickz,Reshetnyak,selinger,lopatina,Rzoska}.
The inclusion of colloidal particles in a LC matrix may produce a distortion of the nematic elastic field giving rise to long range interactions between the particles \cite{yoshida,eva,TGB}. These elastic interactions depend on the size, and shape of the particles, the anchoring condition at the LC--NP interfaces, and the elastic constants of the nematic crystal. In general, nanoparticles could enforce topological defects in LC medium in order to accommodate elastic distortions.

 In the present paper, we use the PRS technique developed by JCPP, and birefringence measurements to investigate the impact of spherical NP, surface treated to give homeotropic anchoring, on the nematic order parameters of a mesogene that exhibits a nematic phase of wide temperature range. We measure the nematic orientational order as function of the NP concentration, and the temperature. Our results on $\left\langle P_2\right\rangle$ and  $\left\langle P_4\right\rangle$ show a strong dependence of nematic order on the NP concentration.

 The remaining of this paper is organised as follows. Section II is devoted to the materials and experimental techniques. In Section III, we review the principle of the Polarised Raman Scattering method we used \cite{Jen}.  In Section IV, are presented some experimental results and their analysis. Section V, is devoted to discussion, and a qualitative description of the experimental results. In the final Section VI, are given some conclusions.

\section{materials and experimental techniques}

In the present investigation, we used the liquid crystal compound 4-n-pentyloxyphenyl-4'-n-octyloxybenzoate (5OO8) with molecular formula  $C_{26}H_{36}O_4$ and molecular weight $MW =412.562$ g/mol. 5OO8 shows the following phase transition sequence when cooling from the isotropic (I) phase: I--85.7$^{\circ}$C--N--64.3$^{\circ}$C--SmA-62.1$^{\circ}$C--SmC-45$^{\circ}$C--Cr. In heating, the SmC phase does not appear, that is, $5OO8$ exhibits a monotropic SmC phase.

Core-shell quantum dots ($QD$) composed of a $CdSe$ spherical core, with diameter of 6.7nm, capped with epitaxial $ZnS$ shell, of thickness 0.6nm, were purchased by PlasmaChem. The molar weight of the core is 671 kDa. The surface hydrophobic layer consists of mostly trioctylphosphine oxide ($TOPO$) that is an organophosphorus compound with the formula $OP(C_8H_{17})_3$.  Since a $TOPO$ molecule has approximately a length of $0.7$nm \cite{topo},  $TOPO$-coated quantum dots have approximately a total diameter of 9.3nm.

Several mixtures of the $LC$ and $QDs$ were prepared with the following mixing protocol. After the $QDs$ were dispersed in toluene, the solution was sonicated for 1h. The mixtures of $QDs$ with the $LC$ were prepared by solving the $LC$ in toluene and adding in the solution, a known volume of the $QDs$ dispersion. Before the evaporation of the solvent while stirring with a magnet, the mixture was sonicated for 2h, at least. The per weight concentration of the mixtures was $\chi=$ 0.001, 0.0025, 0.0035, 0.004, and 0.01, called hereafter $M_1,M_2,M_3,M_4,M_5$ respectively. $\chi$ is defined as the fraction of the $QDs$ mass $m_{QD}$ over the total mass of the mixture, that is, $\chi=m_{QD}/m$ where $m=m_{LC}+m_{QD}$. For each concentration some planar and homeotropic cells were prepared  in order to be used for polarized optical microscopy and Raman measurements. In particular, we used planar cells from Instec, with a polyimide coating. The thicknesses of the cells were 9$\mu$m and 20$\mu$m (Instec in Boulder Colorado).
We also used homeotropic cells from Instec with a homeotropic polyimide coating. The thicknesses of homeotropic cells were 9$\mu$m and 20$\mu$m. Finally, the $LC-NP$ mixture was filled in the cell gap via capillary forces close to the isotropic phase ($T\simeq T_{NI}+5\,\mathrm{K}$).

\subsection{Polarized Opical Microscopy (POM)}

The quality of the alignment, and dispersion; and eventually the textures in the samples were observed under a Leica $DM2500P$ polarizing optical microscope in the transmission mode. The birefringence of the $LC$ compound as function of temperature was measured using a Berek compensator in a planar geometry cell of known thickness. The thickness of the cell was measured by an interferential method. In particular, the heating and cooling rates were of the order of 0.1-0.4 K min$^{-1}$ in the vicinity of the $I-N$ and $N-SmA$ phase transitions. The refractive index of the compound was measured by means of spectroscopic methods ($\theta$--metrisis).

\subsection{Polarised Raman Spectroscopy (PRS)}

Polarized Raman spectra were  acquired in a backscattering geometry, along the axis perpendicular to the substrate plates, using a micro-Raman system with a Jobin Yvon  T64000 triple monochromator including a liquid nitrogen cooled charge-coupled-device (CCD) detector. The 514.5 nm line of an Ar-ion laser was used as an excitation source operating at 5mW. The resolution was set to  3cm$^{-1}$. The polarized Raman spectra were measured  using a  40 times magnification dry objective with numerical aperture 0.4. The temperature of the sample was controlled by a Linkam (Linkam Scientific Instruments THMS 600) heating stage with an accuracy of 0.1K. The measurements as a function of temperature were made over the whole nematic phase. Each spectra set was recorded during 12-30 min depending on the signal quality. The latter was worst for the homeotropic geometry. The experimental set-up is schematically
sketched in Figure 1. The quality of the samples was always tested by POM before using them at the micro-Raman set-up.

\begin{figure}
\includegraphics[scale=0.5]{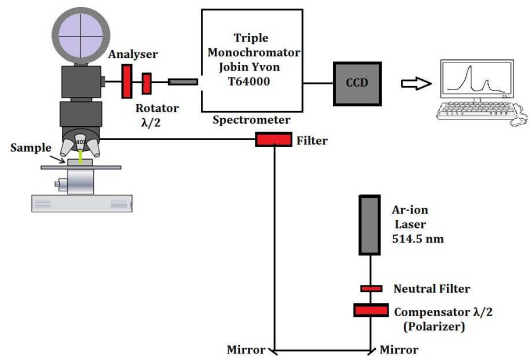}
\caption{Experimental set-up for Raman Polarization Spectroscopy.}
\label{fig1}
\end{figure}




\section{Polarised Raman Scattering}

In this Section, we review the principles of the method \cite{Jen,deJeu,Miyano}.
The long range orientational order is described by means of the orientational
order parameters ($OOP$) which specify the orientational distribution of the long molecular axis around the director
$\mathbf{n}$. In the simplest case, the molecules are assumed to possess an effective cylindrical symmetry, and the orientational distribution function $f(\beta)$  depends only on the angle $\beta$ between the molecular long axis and the director $\mathbf{n}$. For our samples, the z-axis of the laboratory frame is taken parallel to $\mathbf{n}$, that is, z-axis is perpendicular to the plates composing the cell for homeotropic samples and parallel to the plates for planar samples. In this case, $f(\beta)$ can be expanded in terms of Legendre polynomials, $P_L(\cos\beta)$, with coefficients proportional to the statistical averages, $\left\langle P_L(\cos\beta)\right\rangle$, that is to the corresponding OOP. The first two
statistical averages $\left\langle P_2\right\rangle$  and $\left\langle P_4\right\rangle$ can be measured using
various different experimental methods, including polarized
Raman spectroscopy, and x-ray diffraction.

Raman scattering is an inelastic scattering process that arises as a result of the interaction of light with the derivatives
of the second rank polarisability tensor $\alpha_{ij}$, with respect to the
distortion coordinates $q_k$ in a physical medium. The subscripts $i,j$ refer to the polarisation of the scattered and the incident light respectively.
The integrated intensity of the scattered Raman light is proportional to the square of the polarisability derivative with respect to $q_k$, that is\begin{eqnarray}
I_s \propto\left( \dfrac{\partial \alpha}{\partial q_k}\right)^2_{q_k=0}=(\alpha')^2.
\end{eqnarray} Since the polarisability is a tensor, $I_s$ is written as\begin{equation}
I_s=I_0\left\langle \left(\mathbf{e}_s\,\mathbf{R}\,\mathbf{e}_i \right)^2 \right\rangle
\end{equation}
where  $I_0$ is the incident light intensity, $\mathbf{e}_s$, $\mathbf{e}_i$ are the unit vectors given the direction of the polarisation for the scattered and the incident light respectively, angle brackets denote average about the orientational distribution of the scatters over the Raman scattering volume, and $\mathbf{R}$ is the effective molecular Raman tensor
\begin{eqnarray}
R_m\thicksim\left( \begin{matrix}
a & 0 & 0 \\
0 & b & 0 \\
0 & 0 & 1
\end{matrix} \right)
\end{eqnarray}
where the 3-axis of $\mathbf{R}$ is defined along the symmetry axis of the molecular bond stretch vibration, that is, in our case along the long molecular axis. In the laboratory frame the Raman tensor has the general form

\begin{eqnarray}
R_L=\left( \begin{matrix}
\alpha_{xx}' & \alpha_{xy}' & \alpha_{xz}' \\
\alpha_{yx}' & \alpha_{yy}' & \alpha_{yz}' \\
\alpha_{zx}' & \alpha_{zy}' & \alpha_{zz}'
\end{matrix} \right)
\end{eqnarray}

Experimentally, in backscattering geometry, one needs an homeotropic and a planar cell in order to measure the four independent components of the differential polarisability tensor that are obtained from the following Raman Depolarisation Ratios (RDRs)

\begin{eqnarray}
R_1=C_n\dfrac{\langle \alpha_{yz}'^2\rangle}{\langle \alpha_{zz}'^2\rangle};\,R_2=C_n^{-1}\dfrac{\langle \alpha_{zy}'^2\rangle}{\langle \alpha_{yy}'^2\rangle};\,R_3=\dfrac{\langle \alpha_{yx}'^2\rangle}{\langle \alpha_{xx}'^2\rangle}\,.
\end{eqnarray}
where
\begin{align}
C_{n}=\left( \frac{n_{g}+n_{e}}{n_{g}+n_{o}}\right)^{2}
\end{align}
is a correction factor for the birefringence of the liquid crystal and the LC--glass interface. $n_{g}$ is the refractive index of the fused quartz cell that limits the sample and $n_{o},\,n_{e}$ are respectively the ordinary and the extraordinary refractive indices of the liquid crystal.

For vibration direction parallel to the principal molecular axis of symmetry which forms an angle $\beta$ with the z-axis of the laboratory frame, JCPP \cite{Jen} showed that polarized Raman spectroscopy (PRS) can be used to obtain the first two orientational order parameters, $<P_2>$ and  $<P_4>$, of the angular distribution function, by means of the following equation system
\small{
\begin{align}
\label{dptop1}
\dfrac{<{\alpha'}^2_{xx}>}{A^2} &= \frac{1}{9}+\frac{3B}{16}+\frac{C}{4}+\frac{D}{18}+\frac{11D^{2}}{288}+(\frac{B}{8}+
\frac{C}{2}-\frac{D}{6}-\frac{5D^{2}}{48})\\&<cos^{2}\beta>
+(\frac{3B}{16}-\frac{3C}{4}+\frac{3D^{2}}{32})<cos^{4}\beta> \notag\\
\label{dptop2}
 \dfrac{<{\alpha'}^2_{xy}>}{A^2} &= \frac{B}{16}+\frac{C}{4}+\frac{D^{2}}{32}+(\frac{3B}{8}-\frac{D^{2}}{16})<cos^{2}\beta>\\
&+(\frac{B}{16}-\frac{C}{4}+\frac{D^{2}}{32})<cos^{4}\beta> \notag \\
\label{dptop3}
\dfrac{<{\alpha'}^2_{xz}> }{A^2} &= \frac{B}{4}+\frac{C}{4}-(\frac{3C}{4}-\frac{D^{2}}{8})<cos^{2}\beta>\\
&-(\frac{B}{4}-C+\frac{D^{2}}{8})<cos^{4}\beta> \notag \\
\label{dptop4}
\dfrac{<{\alpha'}^2_{zz}>}{A^2} &= \frac{1}{9}+\frac{B}{2}-\frac{D}{9}+\frac{D^{2}}{36}-(B-2C-\frac{D}{3}+\frac{D^{2}}{6})<cos^{2}\beta>\\
&+(\frac{B}{2}-2C+\frac{D^{2}}{4})<cos^{4}\beta> \notag
\end{align}}

where
\begin{align}
&A = a+b+1\\
&B = \frac{\frac{1}{4}(a-b)^{2}}{A^{2}}\\
&C = 0\\
&D = \frac{2-a-b}{A}\,.
\end{align}
The full expressions in the general case of $A,B,C,D$ are given in \cite{Jen}.
In the isotropic phase, both $\left\langle P_2\right\rangle$ and $\left\langle P_4\right\rangle$ are
zero and the depolarization ratio $R_{iso}$  is written as
\begin{align}
R_{iso}=\dfrac{3(1-a-b-ab+a^2+b^2)}{5(a+b+1)^2+4(1-a-b-ab+a^2+b^2)}\,
\end{align}
that for vibrations with uniaxial symmetry, $a=b=r$, simplifies to \begin{align}
R_{iso}=\dfrac{(1-r)^2}{3+4r+8r^2}
\end{align}

If one takes $r$ from the above equations and keeps it constant then  $<P_4>$ is abnormally low or even negative. If $r$ is treated as a fitting parameter then the obtained values are in good agreement with the theory. Some authors use $a\ne b$ and perform fitting \cite{Miyano}. The quality of the fitting is controlled by computing $R_3$ and comparing it with the experimental one.  To avoid fitting procedure one has to solve analytically the above set of equations  (\ref{dptop1} -\ref{dptop4}).
Inspection of the above equations shows that one has four experimental quantities from which can be analytically calculated the four unknowns, that is, the derived molecular polarisabilities $a$ and $b$ and the mean values $<cos^{2}(\beta)>$ and $<cos^{4}(\beta)>$ and consequently the uniaxial order parameters $<P_2>$ and $<P_4>$. The analysis of our data has been performed both by fitting procedure using matlab software, and  using the analytical solution of the system equations (\ref{dptop1} -\ref{dptop4}) obtained with the aid of mathematica software. In our case, both methods give similar results.


\section{Results}

Samples of four different compositions were studied, the pure $LC$--compound $5OO8$ and five mixtures with $\chi=0.001, 0.0025, 0.0035,0.004, 0.01$ per wt in $QDs$. For each composition a planar and an homeotropic cell were used to acquire the six combinations of the polarization required  by the JCPP method. In all cases the measurement procedure started deep in the isotropic phase at  $\Delta T\simeq T_{NI}-T=-5\,\mathrm{K}$, where, $R_{iso}$ was measured for all  samples, that is, for planar and homeotropic cells and for all mixtures. The measured, in the mean, value $R_{iso}=0.44\pm 0.02$, was calculated using the integrated intensities of the Raman lines and its value was essentially the same, within experimental errors, for all mixtures and the pure LC compound. This value is higher than the usually measured values in most $LCs$ systems \cite{Jen,Miyano}. Then the sample was cooled down slowly in the nematic phase at a rate of 1K/min. Once the desired temperature was reached and $T$ was stabilized,  the Raman spectra were acquired. The measurements were done over the whole nematic range in steps of $5\,\mathrm{K}$.

\begin{figure}
\includegraphics[scale=0.6]{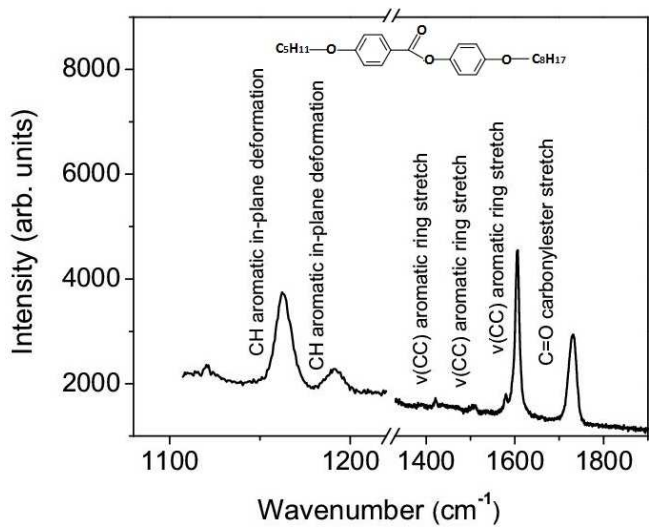}
\caption{Raman spectrum of the pure $5OO8$ $LC$--compound, in the nematic phase at $\Delta T=T_{NI}-T=15.7\,\mathrm{K}$, with peak assignment.}
\label{fig2}
\end{figure}

Figure 2, shows a typical Raman spectrum of the $5OO8$ pure $LC$--compound in the nematic phase and the peak assignment.
For liquid crystal systems the most commonly used Raman line is the uniaxial $C-C$ stretching mode scattering (phenyl breathing mode) of the two phenyl rings with a Raman shift of 1607 cm$^{-1}$, which is strongly polarised along the long molecular axis and is well isolated from other lines, as can be seen in Figure 2 for the present compound. For these reasons, we chose this mode for the calculation of the DPRs in our investigations.

\begin{figure}
\includegraphics[scale=0.35]{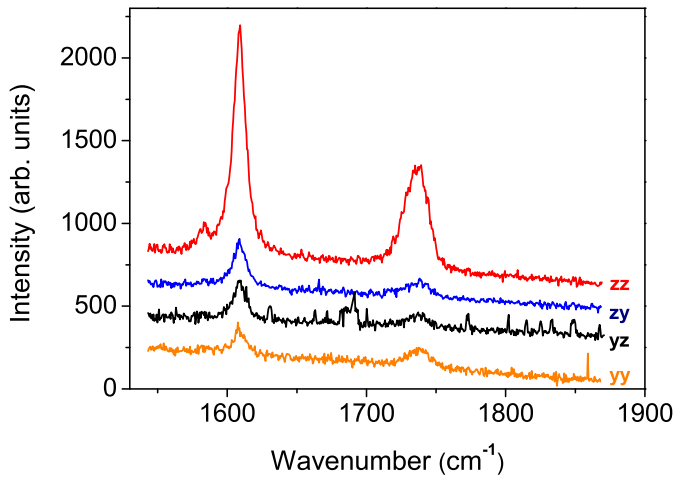}
\caption{Polarized components of the Raman scattering spectra of the $5OO8$ $LC$--compound at $\Delta T=5.7\,\mathrm{K}$, in the nematic phase, in planar geometry.}
\label{fig3}
\end{figure}

\subsection{pure 5OO8}

Figure 3, shows typical PRS-spectra of the liquid crystalline compound $5OO8$ at
$\Delta T=5.7\,\mathrm{K}$ in a planar cell for the polarisations $I_{yy}$, $I_{zz}$, $I_{yz}$ and $I_{zy}$. The measured DPRs as function of temperature are listed in Table-\ref{dpr0}. $R_1$ is an increasing function of the temperature while $R_2$ is a decreasing function of the temperature.

\begin{figure}
\includegraphics[scale=0.35]{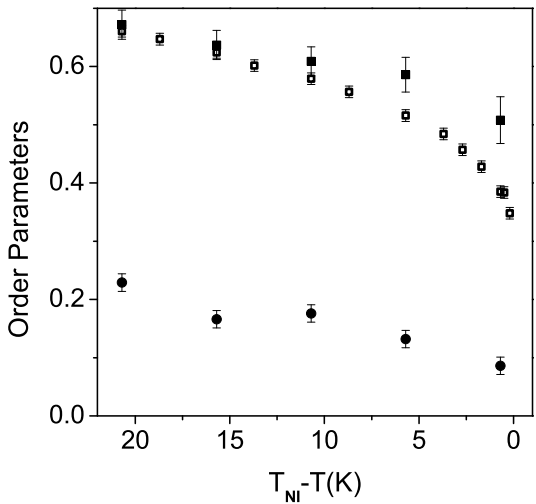}
\caption{$<P_2>$, $<P_4>$  orientational order parameters vs temperature of the LC--compound $5OO8$ in the nematic phase. Black solid squares: $<P_2>$ from Raman measurements, open squares $<P_2>$ from  birefringence measurements, black points: $<P_4>$ from Raman measurements. }
\label{fig4}
\end{figure}
Figure 4 shows the order parameters $<P_2>$ (solid squares) and $<P_4>$ (solid points) as function of the temperature in the nematic phase. Note that the obtained $<P_4>$ is always positive and monotonic. In the same figure is given $<P_2>$ calculated from optical birefringence measurements (open squares) by means of a tilting compensator.
The numerical values of $<P_2>$, $<P_4>$, and of the elements $a$,$b$ of the derived polarizability Raman tensor are listed in
Table-\ref{top}. $a$ and $b$ are both monotonic functions of temperature.
If one supposes cylindrical symmetry, that is, $a=b=r$ then $R_{iso}=0.44$ that yields $r=-0.09$. The calculated values of $<P_2>$, and $<P_4>$ are essentially the same as previously if a temperature dependance of $r$ is assumed (data not shown). The overall agrement between the two experimental methods in what concerns $<P_2>$ is good. Birefringence results to slightly lower values of  $<P_2>$ especially close to $T_{NI}$ where director fluctuations are stronger than deeper in the nematic phase. As Raman measurements are sensitive to the core part of the molecules they are less sensitive to fluctuations than birefringence.


\begin{table}
	\caption{Temperature dependence of the Raman depolarization ratios for the pure nematic liquid crystal 5OO8.} \label{dpr0}
 \centering
		\begin{tabular}{cccc}
			\hline
			$\Delta T\,[\mathrm{K}]$& $R_{1}$ & $R_{2}$ & $R_{3}$\\
			\hline\hline
			20.7 & 0.141 & 3.058 & 0.671\\
			\hline
			15.7& 0.161 & 2.887 & 0.660\\
			\hline
			10.7 & 0.162 & 2.043 & 0.682 \\
			\hline
			5.7 &0.177 & 1.937 & 0.641\\
			\hline
			0.7 &0.201& 1.403& 0.592 \\
			\hline
		\end{tabular}
\end{table}

\subsection{mixtures 5OO8 and NPs}

\begin{figure}
\includegraphics[scale=0.38]{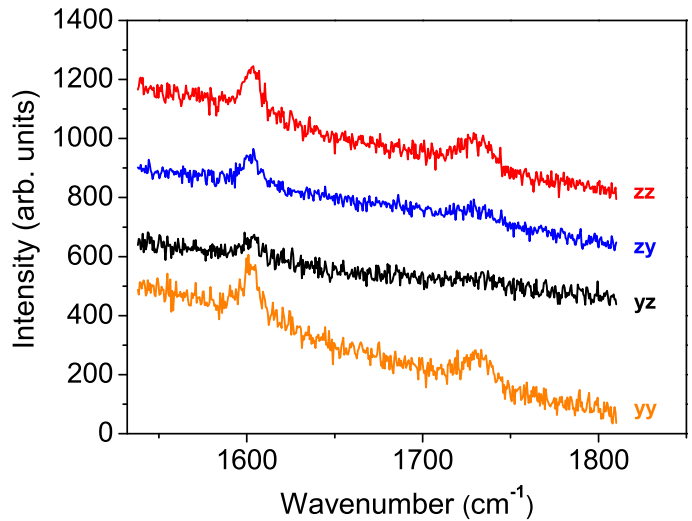}
\caption{Raman polarisation spectra of the $M_4$, $\chi=0.4\%$ wt, mixture in the nematic phase, at $\Delta T=-4.7$K.}
\label{fig6}
\end{figure}
For the mixtures, the Raman band intensity varies strongly with the $NP$ concentration. Typical Raman intensity profiles for four polarization configurations are shown in Figure 5, for the case of the mixture $M_4$ at $\Delta T=-4.7\,\mathrm{K}$ in planar geometry. It is obvious that the influence of the $NP$ on the order parameter is of some importance.

The measured depolarisation ratios $R_1$, $R_2$, and $R_3$ as function of temperature and for all mixtures, are given in Tables \ref{dpr01}-\ref{dpr10}. For the mixture of the lowest concentration in $NPs$, namely $M_1$, the values of the DPRs are almost the same as those of the pure $LCs$. In the contrary, for $M_3$, $M_4$ and $M_5$, $R_1$ and $R_2$ change significantly in respect to the corresponding DPR of the pure $5OO8$. In particular $R_1$ and $R_2$ are strongly affected by the presence of the $NP$ for mixtures $M_4$ and $M_5$. Note that the absolute difference  $|R_1-R_2|$ is a decreasing function of $\chi$, signaling a decrease of the $OP$.
The measured DPRs indicate that the impact of the $NPs$ is weak on $R_3$, while $R_{iso}$ is practically unaffected since its measured variations are in the range of the experimental error.

\begin{table}
	\caption{Temperature dependence of the Raman depolarization ratios for the pure nematic liquid crystal 5OO8 + 0.1\% wt CdSe-ZnS nanoparticles.} \label{dpr01}
	\begin{center}
		\begin{tabular}{cccc}
			\hline
			$\Delta T\,[\mathrm{K}]$ & $R_{1}$ & $R_{2}$ & $R_{3}$\\
			\hline\hline
			20.6 & 0.117 & 2.361 & 0.656\\
			\hline
			15.6& 0.137&1.841 & 0.640\\
			\hline
			10.6 & 0.186 & 1.740& 0.618 \\
			\hline
			5.6&0.201& 1.660 & 0.654\\
			\hline
			1.6&0.215& 1.462& 0.560 \\
			\hline
		\end{tabular}
	\end{center}
\end{table}

\begin{table}
	\caption{Temperature dependence of the Raman depolarization ratios for the pure nematic liquid crystal 5OO8+ 0.25\% wt CdSe-ZnS nanoparticles.} \label{dpr025}
	\begin{center}
		\begin{tabular}{cccc}
		 \hline
		  $\Delta T\,[\mathrm{K}]$& $R_{1}$ & $R_{2}$ & $R_{3}$\\
		 \hline\hline
		 20.4& 0.143 & 2.238 & 0.531\\
		 \hline
		 15.4& 0.154&1.908 & 0.518\\
		 \hline
		 10.4& 0.161 & 1.880& 0.505\\
		 \hline
		 5.4&0.170& 1.695& 0.524\\
		 \hline
		 0.4&0.304& 0.888& 0.502 \\
		 \hline
		 0.2&0.334&0.670&0.551 \\
		 \hline
		\end{tabular}
	\end{center}
\end{table}

\begin{table}
	\caption{Temperature dependence of the Raman depolarization ratios for the pure nematic liquid crystal 5OO8+ 0.35\% wt CdSe-ZnS nanoparticles.} \label{dpr035}
	\begin{center}
		\begin{tabular}{cccc}
			\hline
			$\Delta T\,[\mathrm{K}]$& $R_{1}$ & $R_{2}$ & $R_{3}$\\
		 \hline\hline
		 18& 0.258& 1.756& 0.600\\
		 \hline
		 15& 0.246 & 1.542 & 0.534\\
		 \hline
		 10& 0.255 &1.467& 0.535 \\
		 \hline
		 5&0.244 & 1.286 & 0.503\\
		 \hline
		 1&0.343& 0.687& 0.558 \\
		 \hline
		\end{tabular}
	\end{center}
\end{table}

\begin{table}
	\caption{Temperature dependence of the Raman depolarization ratios for the pure nematic liquid crystal 5OO8 + 0.4\% wt CdSe-ZnS nanoparticles.} \label{dpr05}
	\begin{center}
		\begin{tabular}{cccc}
			\hline
			$\Delta T\,[\mathrm{K}]$& $R_{1}$ & $R_{2}$ & $R_{3}$\\
			\hline\hline
			19.7 & 0.279 & 0.671 & 0.580\\
			\hline
			14.7 & 0.360 & 0.586 & 0.570\\
			\hline
			9.7 & 0.430 &0.659 & 0.530 \\
			\hline
			4.7&0.433 & 0.551 & 0.510\\
			\hline
			0.7&0.480& 0.529& 0.520 \\
			\hline
		\end{tabular}
	\end{center}
\end{table}

\begin{table}
	\caption{Temperature dependence of the Raman depolarization ratios for the pure nematic liquid crystal 5OO8+ 1\% wt CdSe-ZnS nanoparticles.} \label{dpr10}
	\begin{center}
		\begin{tabular}{cccc}
			\hline
			$\Delta T\,[\mathrm{K}]$& $R_{1}$ & $R_{2}$ & $R_{3}$\\
			\hline\hline
			19.3 & 0.461 & 0.778 & 0.598\\
			\hline
			14.3& 0.440 &0.704 & 0.585\\
			\hline
			9.3 & 0.448& 0.544 & 0.570 \\
			\hline
			4.3&0.574 & 0.540 & 0.549\\
			\hline
			0.3&0.460& 0.490& 0.450 \\
			\hline
		\end{tabular}
	\end{center}
\end{table}

Figures 6 and 7, show the calculated $<P_2>$ (solid symbols), from the experimental $DPRs$, as function of the temperature and mixture concentration. The values of  $<P_2>$ obtained by birefringence measurements are included for comparison (corresponding open symbols). Both experimental methods give similar results for $<P_2>$. Note that the assumption of cylindrical symmetry $a=b=r$ (not presented here) does the analysis inconsistent with the experimental measurements. The calculated values of $a,\,b,\,<\cos^{2}(\beta)>,\,<\cos^{4}(\beta)>,\,<P_{2}>,\, \& \,<P_{4}>$ as function of temperature and for all mixtures, are given in Tables \ref{top01}-\ref{top10}.

For low $NPs$ concentrations (Fig.6), up to $\chi=0.001$, the  dependence of $<P_2>$ on $\chi$ does not result to any appreciable change of $<P_2>$. Nevertheless, the variation of $<P_2>$ becomes steeper with $T$, and the first order $I-N$ transition becomes softer. At low temperatures $<P_2>$ becomes slightly stronger than its value in the pure compound. $<P_4>$ remains positive but the amplitude of its variation increases from 0.14 to 0.26 indicating a stronger dispersion of the molecular distribution about the local nematic director. For $\chi=0.0025$ (Fig.6),  $<P_2>$ becomes a little weaker than in pure $5OO8$. At $\chi=0.0035$ (see Fig.7),  $<P_2>$ decreases further. When $\chi$ increases at about 0.004 a strong decrease of the nematic order is measured while for higher values, up to 0.01, essentially no further destruction of the order parameter is observed. For larger values of $\chi$ our samples present strong  phase separation effects and therefore we did not attempt to perform measurements.
\begin{figure}
\includegraphics[scale=0.4]{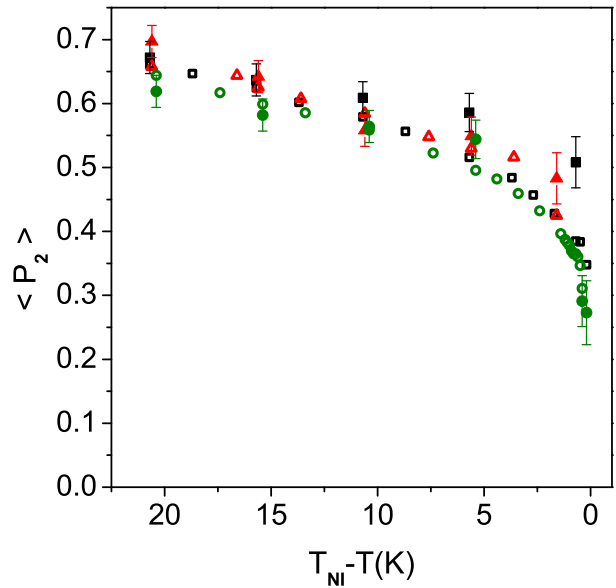}
\caption{Temperature dependence of $\left\langle P_2\right\rangle$ vs temperature for low concentrations ($\chi < 0.003$), in the nematic phase. Solid symbols for Raman data. Pure 5OO8: black solid squares; $M_1$: red solid triangles;  $M_2$: green bullets. Open symbols concern birefringence measurements. Pure 5OO8: black open squares; $M_1$: red open triangles;  $M_2$: green circles.}
\label{fig6}
\end{figure}

\begin{figure}
\includegraphics[scale=0.4]{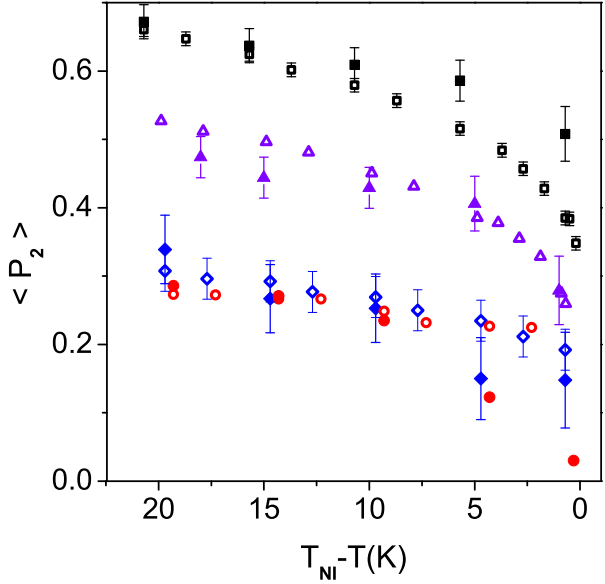}
\caption{Temperature dependence of $\left\langle P_2\right\rangle$ vs temperature in the nematic phase for $\chi> 0.003$. Solid symbols for Raman data. Pure 5OO8: black solid squares; $M_3$: violet solid triangles;  $M_4$: blue solid diamonds;  $M_5$: red bullets. Open symbols concern birefringence measurements. Pure 5OO8: black squares; $M_3$: violet open triangles;  $M_4$: blue open diamonds;  $M_5$: red circles.}
\label{fig7}
\end{figure}

Apparently the effect of $NPs$ at low enough concentration is not destructive for the nematic tensorial order parameter but above a critical concentration, or range of concentrations, the nematic order parameter is strongly affected by the presence of $NPs$, and eventually results to a structural transition for the $NP$ that should be investigated by SAXS measurements. What we can confirm from our experimental data is the presence of low nematic order for $\chi\gtrsim 0.004$, observed by $POM$. Probably the system enters a crossover regime towards a distorted nematic phase of low order parameter (like a paranematic phase) as it is also suggested from the data in Fig.7.

\section{\textbf{Discussion}}

The $NPs$ are spherical and functionalised with $TOPO$ that favors homeotropic alignment at their surface. In an oriented $LC$ cell, in general, one expects a deformation of the nematic orientation around the $NPs$. Whether the anchoring of the director at the surface of the $NP$ actually leads to the deformation of the director field depends on the elastic properties
of the system, and the diameter $D$ of the $NPs$. Let $L_{ext}=K/W$ be the surface extrapolation length\cite{degennes,samo}, where $K$ is a mean elastic constant of the nematic, and $W$ the anchoring energy. If $D<<L_{ext}$ the director field is weakly perturbed by the presence of the particles, while if $D>>L_{ext}$ the opposite is true.
This might lead to the emergence of different kind of topological defects in the nematic phase, depending on the anchoring conditions, shape and size of the particle. Note that particles have in general different impact on topological defects. If they are small enough (comparable to the nematic correlation length \cite{LND}) and if their locally enforced $LC$ structure is compatible with the core structure of a defect, then they tend to assemble in cores of defects. In these cases they might stabilize topological defects if they are enforced (e.g., by chirality  in Blue phases \cite{yoshida,eva,graphBP} and Twist Grain Boundary phases \cite{TGB}) to a system. On the other hand, they might give rise to additional defects if they effectively enforce a local structure similar to a topological defect due to the topological charge conservation law.

Our samples are monodomain crystals with no macroscopic disclination lines in the volume or at the surfaces as has been verified in $POM$. In this case, $NPs$ could partially aggregate (coacervation) giving rise to regions that are diluted in respect to the average concentration of an homogeneous sample and regions of higher concentrations. In fact, such regions are present in some samples and their number and size increases strongly for $\chi>>0.01$. In our micro-Raman experiment, we paid particular attention to perform  measurement away from regions with agglomerates, if any. In case that a few agglomerates were present we performed measurements at two/three different points of the sample. In case, the obtained results were significantly different we changed the sample.

Our experimental data suggest the existence of a crossover between nematic and distorted nematic regimes, on increasing the concentration of nanoparticles. Below, we estimate the critical condition for this crossover. We express the free energy of the system as

\begin{equation}
F=\int f_{e}d^{3}\mathbf{r}+\int f_{a}d^{2}\mathbf{r}.  \label{F}
\end{equation}
The first integral is carried over the $LC$ body, the second one over the
$LC$-nanoparticle interfaces, $f_{e}$ stands for the elastic free energy
density, and $f_{a}$ is the free energy anchoring contribution at the
interfaces. Experiments suggest that disclinations are not present.
Consequently, we neglected the condensation free energy penalty in Eq.(\ref{F}) and assume that the degree of nematic uniaxial ordering $<P_{2}>$ is
approximately constant over the sample. The observed crossover
results from the competition between the contradicting tendencies of $f_{e}$
and $f_{a}$. The elastic term enforces spatially homogeneous nematic
ordering. Using the single elastic constant approximation we express $f_{e}$ as

\begin{equation}
f_{e}=\frac{K}{2}\left\vert \nabla \mathbf{n}\right\vert ^{2}.  \label{fe}
\end{equation}%
Here $K$ is the representative positive Frank elastic constant and $\nabla $
is the gradient operator. We model the interface contribution using the
classical Rapini-Papoular-type approximation

\begin{equation}
f_{a}=W\left( 1-\left( \mathbf{n}\cdot \mathbf{v}\right) ^{2}\right) .
\label{fa}
\end{equation}
The positive anchoring strength constant $W$ locally favors alignment of $%
\mathbf{n}$ along the $LC-NP$ interface surface normal $\mathbf{v}$,
corresponding to the homeotropic anchoring condition.

To estimate the  conditions for the \textbf{crossover} we assume
that before the transition the nematic structure is essentially spatially
homogeneous. Entering the crossover range the nematic director field
becomes spatially distorted. We henceforth refer to these competing
configurations as the \textit{homogeneous} (HOM) and \textit{distorted }(DIS)
structure, respectively.

In a rough approximation we set that in the \textit{homogeneous} structure
$LC$ is homogeneously aligned along a single symmetry breaking direction
\textbf{n}. It holds $f_{e}=0$ and the free energy penalties arise only at
$NP-LC$ interfaces. The corresponding total free energy cost $F_{HOM}$ is
approximately given by

\begin{equation}
F_{HOM}\sim \frac{W}{2}N_{NP}a_{NP},  \label{FH}
\end{equation}%
where $a_{NP}=\pi D^{2}$ is the surface area of a nanoparticle.

In the \textit{distorted} structure we set that homeotropic anchoring is
obeyed, i.e. $f_{a}=0$. Furthermore, we assume that the \textit{distorted}
nematic pattern is characterized by a typical length $\xi _{n}$, hence $
f_{e}\sim \frac{K}{2\xi _{n}^{2}}$. It follows

\begin{equation}
F_{DIS}\sim \frac{KV}{2\xi _{n}^{2}},  \label{FD}
\end{equation}
where $V$ stands for the sample volume.

The critical condition for the $HOM-DIS$ crossover is estimated from the condition $F_{DIS}=F_{HOM}$. It follows

\begin{equation}
\phi _{c}\sim \frac{D L_{ext}}{6\xi _{n}^{2}},  \label{fc}
\end{equation}%
where $\phi _{c}$ stands for the critical volume concentration. The volume concentration of $NPs$ is defined as

\begin{equation}
\phi =\frac{N_{NP}v_{NP}}{V}  \label{f}
\end{equation}%
where $N_{NP}$ stands for the number of nanoparticles and $v_{NP}=\pi D^{3}/6$ is the volume of a spherical nanoparticle of diameter $D$. In diluted regime it holds $\phi \simeq \chi \rho _{LC}/\rho _{NP}\simeq \chi/6$, where $\rho _{LC}$ and $\rho_{NP}$ are mass densities of $LC$ and $NPs$, respectively.
Note that $\xi _{n}$ depends on the concentration of $NPs$. In case of
spatially homogeneous distribution of $NPs$ the average separation $d_{NP}$
between neighboring $NPs$ is given by

\begin{equation}
d_{NP}\sim \left( \frac{\pi }{6}\right)^{1/3}\frac{D}{\phi^{1/3}}.
\label{dNP}
\end{equation}
For example, for $D\sim 9$ nm and $\phi =0.001$ it holds $d_{NP}$ $\sim 73$
nm. Since at the critical condition $d_{NP}$ sets an upper limit for $\xi_{n}$ and therefore to $L_{ext}$, one can test the condition $D/L_{ext}>>1$, that is, if the $NPs$ deform the $LC$-host, by setting $d_{NP}\sim \xi_{n}$. It follows

\begin{equation}
\phi_{c}\sim \left( \frac{L_{ext}}{D}\right)^{3}\frac{1}{6\pi^{2}}.
\label{fc1}
\end{equation}

Figure 8 suggests that the system enters a crossover regime between $\chi
=0.0035$ and $\chi =0.004$. In our estimate we set $\chi \sim 0.0038$.
Since $\rho_{NP}/\rho_{LC}\simeq 6$ we obtain $\phi_{c}\sim 0.0006$.
Taking into account Eq.(\ref{fc1}) it follows $L_{ext}\sim 3.1$
nm. Therefore, $D/L_{ext}\sim 3$, which is consistent with our assumptions.
Namely, for an isolated nanoparticle the anchoring strength $W$ is
sufficiently strong to overwhelm elastic forces providing $D/L_{ext}>1$. For
$K\sim 10^{-12}$ J/m we obtain $W\sim 3\,10^{-4}$ J/m$^{2}$,
corresponding to a strong but reasonable anchoring strength value \cite{lev}.

In the past \cite{li,Reznikov,freederickz}, order parameter and dielectric properties of hybrid $NP+LC$ systems have been investigated in the case of ferroelectric $NPs$ with dimension greater than 50 nm. A strong enhancement of the orientational order was observed while the clearing temperature increased up to 40K compared to the pure $LC$-host. These effects arrives, according to \cite{li}, from the strong effective electric field due to the particles. However, there are limitations related to the size of the ferroelectric $NPs$ in order to exhibit ferroelectric behaviour. In our case the $NP$ act as disorder sources inducing a decrease of the nematic order and a smooth decrease ($\sim 1\,K$) of the clearing temperature. As we qualitatively demonstrated with the above  presented minimal model a strong enough anchoring in combination with the shape incompatibility between the $NPs$ and the nematic direction may result to disorder effects. Recently \cite{Rzoska}, experimental evidence about $NP$ induced disorder has been reported by means of broadband spectroscopy.

\section{Conclusions}
In summary, we have experimentally investigated the influence of spherical shape nanoparticles, dispersed in a nematic host, on the nematic order parameter. On increasing the concentration, $\chi$, of NPs we observe a crossover-type structural change at the cross-over concentration $\chi_c \approx 0.004$, separating two qualitatively different regimes. In the regime $\chi <\chi_c$  samples display roughly pure-LC behavior. Nevertheless, the orientational order varies in a steeper way with the temperature in comparison with the pure-LC sample. On crossing $\chi_c$  the nematic degree ordering exhibits substantial drop with respect to the pure-LC reference sample. On increasing concentration above $\chi_c$ the degree of ordering displays relative weak changes at a given temperature. This suggests a structural transition from a pure-like nematic to a weakly distorted nematic ordering. The latter exhibits long range or quasi-long order and its structural details are of our future interest. The reason behind this conjectural structural transition are orientational frustrations at NP-LC interfaces. Based on our experimental data we estimated the anchoring interaction strength at the interfaces. Of course, a sudden drop in the order parameter above some critical concentration of NPs
could be also due to a phase separation mechanism, which we analyze in the
Appendix. According to this scenario a two phase pattern is formed,
consisting of the so-called \textit{rich} and \textit{depleted} phase (see
the Appendix). If this is the case the measured order parameter would
represent the average response of these regions. The \textit{depleted} phase
is expected to exhibit bulk-like nematic ordering. On the other hand NPs are
expected to influence degree of nematic ordering in the \textit{rich} phase.
In the Appendix we demonstrate that the average response could explain the
observed sudden drop in $\left\langle P_{2}\right\rangle $ above some
critical concentration of NPs providing that the amplitude of nematic
ordering in the \textit{rich} phase is relatively low. However, this is in
contradiction with the assumption that a nanoparticle acts as a local
ordering field, yielding frustration on a larger length scale.  Moreover, we have never observed any phase separation in POM. Therefore, we
believe that the phase separation mechanism is not responsible for the
observed behavior. X-rays experiments are planned to resolve this question.

\section{Acknowledgments}

C.K. acknowledges financial support from the Hellenic Foundation for Research and Innovation for PhD Candidates (No. 1318), funded by the General Secretariat for Research and Technology (GSRT) and the Hellenic Foundation for Research and Innovation (HFRI). S.K. acknowledges the financial support from the Slovenian Research Agency (research core funding No. P1-0099).

\section{Appendix: Phase separation}

In the following we discuss possibility of phase separation and its impact
on average degree of nematic ordering. In our rough estimate we describe the
average degree of LC ordering by the spatially averaged uniaxial order
parameter $s=\left\langle P_{2}\right\rangle $ and the volume concentration
of nanoparticles $\phi $. The corresponding average free energy density is
expressed as $\left\langle f\right\rangle \sim \left\langle
f_{c}\right\rangle +\left\langle f_{e}\right\rangle +\left\langle
f_{a}\right\rangle +\left\langle f_{m}\right\rangle $, where the
condensation ($\left\langle f_{c}\right\rangle $), elastic ($\left\langle
f_{e}\right\rangle $), NP-LC interface ($\left\langle f_{a}\right\rangle $),
and entropy mixing ($\left\langle f_{m}\right\rangle $) terms are
approximated by

\begin{eqnarray*}
\left\langle f_{c}\right\rangle &\sim &\left( 1-\phi \right) \left(
A_{0}(T-T^{\ast })s^{2}-Bs^{3}+Cs^{4}\right) , \\
\left\langle f_{e}\right\rangle &\sim &\left( 1-\phi \right) \frac{k_{0}s^{2}%
}{\xi _{n}^{2}}, \\
\left\langle f_{a}\right\rangle &\sim &-\phi \left( 1-\phi \right) ws\frac{%
a_{NP}}{v_{NP}}, \\
\left\langle f_{m}\right\rangle &\sim &\frac{k_{B}T}{v_{LC}}\left( 1-\phi
\right) \ln \left( 1-\phi \right) +\frac{k_{B}T}{v_{NP}}\phi \ln \phi
+\kappa \left( 1-\phi \right) \phi .
\end{eqnarray*}

The quantities $A_{0}$, $B$, $C$, are material constants, $T^{\ast }$
describes the supercooling temperature, $k_{0}$ is the bare nematic elastic
constant (i.e., $K\sim k_{0}s^{2}$), $\xi _{n}$ estimates the average linear
scale on which the nematic director field is distorted, $w$ measures the
wetting strength at LC-NP interfaces, $a_{NP}$, $v_{NP}$,$v_{LC}$ stand for
the NP's surface area, NP's volume, and LC molecule's volume, respectively, $%
k_{B}$ is the Boltzmann constant, and $\kappa >0$ stands for the
Flory-Huggins parameter.

Note that in general $T^{\ast }$ is a function of $\phi $. In our simple
modelling we assume that the direct interactions between LC and
nanoparticles are relatively small, suggesting $T^{\ast }\sim T_{0}-\lambda
\phi $. Here $T_{0}$ and $\lambda >0$ are independent of $\phi $ . The
average contribution at LC-NP interfaces is proportional with $\phi \left(
1-\phi \right)$. Namely, this free energy term is absent in limits $\phi
\rightarrow 0$ and $\phi \rightarrow 1$. Furthermore, we assume that NPs
locally favor nematic ordering and consequently $w>0$.

To estimate phase separation tendencies of our system we collect in the
expression for $f$ all the terms proportional with $\phi \left( 1-\phi
\right).$ The corresponding coefficient defines the effective Flory-Huggins
parameter:

\begin{equation*}
\kappa _{eff}=\kappa +A_{0}\lambda s^{2}-\frac{a_{NP}}{v_{NP}}ws.
\end{equation*}

If \ $\kappa _{eff}$ is larger than the critical value $\kappa
_{eff}^{(c)}>0 $ it triggers phase separation. Namely, the contribution of
the effective Flory-Huggins free energy term is minimal for $\phi =0$ and $%
\phi =1$.

Let us suppose that $\kappa _{eff}(s=0)=\kappa <\kappa _{eff}^{(c)}$ in the
isotropic phase. Therefore, the mixture is spatially homogeneous. Below $%
T_{NI}$\ the orientational order appears switching on $s$-dependent
contributions in $s$. Consequently, in presence of ordering the condition $%
\kappa _{eff}>\kappa _{eff}^{(c)}$ could be fulfilled, triggering phase
separation. In the phase separation process two phases are formed, where one
phase is relatively rich in particles in comparison to the 2nd one. One
commonly referees to these phases as the \textit{rich} and \textit{depleted}
phase, which occupy volume $V_{r}$, $V_{d}$, respectively, and $V=V_{r}+V_{d}
$. We characterize the phases by configuration parameters \{$s=s_{r}$ ,$\phi
=\phi _{r}$ \} and \{ $s_{d}$,$\phi _{d}$\}, respectively. It holds

\begin{equation*}
\phi =x\phi _{r}+(1-x)\phi _{d},
\end{equation*}%
where $x=(\phi -\phi _{d})/(\phi _{r}-\phi _{d})$. Note that the phase
separation occurs only within the window $\phi \in ]\phi _{d},\phi _{r}[$.
The average order parameter of the whole sample is then estimated by

\begin{equation*}
s\sim xs_{r}+(1-x)s_{d}.
\end{equation*}

It is expected that the \textit{depleted} phase displays degree of nematic
ordering similar to the bulk nematic phase, which we label by $s_{b}$. On
the contrary, in the \textit{rich} phase ordering could be strongly
influenced by NPs. If one supposes that effectively NP tends to destroy
nematic ordering, we set $s_{d}\sim 0$ and $s_{r}\sim s_{b}$. It follows

\begin{equation*}
s\sim s_{b}\frac{\phi _{r}-\phi }{\phi _{r}-\phi _{d}}<s_{b}.
\end{equation*}

\begin{table}[ht]
\begin{center}
{\small
		\begin{tabular}{ccccccc}
			\hline
			$\Delta T\,[\mathrm{K}]$& a & b & $<\cos^{2}(\beta)>$&$<\cos^{4}(\beta)>$& $<P_{2}>$ & $<P_{4}>$ \\
			\hline\hline
			20.7 & 0.059 & -0.225 & 0.781 & 0.636 & 0.672 & 0.229\\
			\hline
			15.7& 0.067 & -0.232 & 0.758 & 0.602 & 0.637 & 0.166\\
			\hline
			10.7 & 0.107 & -0.263 & 0.739 & 0.588 & 0.609 & 0.176\\
			\hline
			5.7& 0.125 & -0.277& 0.724 & 0.565 & 0.586 & 0.132\\
			\hline
			0.7 & 0.177 & -0.314 & 0.678 & 0.510 & 0.508 & 0.086 \\
			\hline
		\end{tabular}}
	\end{center}\hfill{}
\caption{Temperature dependence of the order parameters of the pure nematic liquid crystal 5OO8.} \label{top}
\end{table}

\begin{table}[ht]
\hfill{}
	\begin{center}
{\small
		\begin{tabular}{ccccccc}
			\hline
			$\Delta T\,[\mathrm{K}]$& a & b & $<\cos^{2}(\beta)>$&$<\cos^{4}(\beta)>$& $<P_{2}>$ & $<P_{4}>$ \\
			\hline\hline
			20.6 & 0.078 & -0.240 & 0.798 & 0.671 & 0.697 & 0.318\\
			\hline
			15.6& 0.113 & -0.268 & 0.761 & 0.624 & 0.642 & 0.251\\
			\hline
			10.6 & 0.130 & -0.280 & 0.705 & 0.544 & 0.558 & 0.111\\
			\hline
			5.6& 0.178 & -0.314 & 0.699 & 0.531 & 0.549 & 0.077\\
			\hline
			1.6& 0.116 & -0.270 & 0.655 & 0.489 & 0.483 & 0.058 \\
			\hline
		\end{tabular}}
	\end{center}
\hfill{}
\caption{Temperature dependence of the order parameters of 5OO8 + 0.1\% wt CdSe-ZnS nanoparticles.} \label{top01}
\end{table}

\begin{table}[ht]
	\hfill{}
	\begin{center}
		{\small
			\begin{tabular}{ccccccc}
				\hline
				$\Delta T\,[\mathrm{K}]$& a & b & $<\cos^{2}(\beta)>$&$<\cos^{4}(\beta)>$& $<P_{2}>$ & $<P_{4}>$ \\
				 \hline\hline
				 20.4& -0.002 & -0.173 & 0.746 & 0.606 & 0.619 & 0.229\\
				 \hline
				 15.4& 0.002 & -0.178 & 0.721 & 0.578& 0.582 & 0.200\\
				 \hline
				 10.4& -0.017& -0.160 & 0.709 & 0.564& 0.564 & 0.184\\
				 \hline
				 5.4& 0.024 & -0.200& 0.696 & 0.548& 0.544 & 0.163\\
				 \hline
				 0.4& 0.165 & -0.306& 0.527 & 0.359 & 0.291& -0.031 \\
				 \hline
				 0.2&0.467&-0.487&0.515&0.343&0.273&-0.056\\
				 \hline
			\end{tabular}}
		\end{center}
		\hfill{}
		\caption{Temperature dependence of the order parameters of 5OO8 + 0.25\% wt CdSe-ZnS nanoparticles.} \label{top025}
	\end{table}

\begin{table}[ht]
	\hfill{}
	\begin{center}
		{\small
			\begin{tabular}{ccccccc}
				\hline
				$\Delta T\,[\mathrm{K}]$& a & b & $<\cos^{2}(\beta)>$&$<\cos^{4}(\beta)>$& $<P_{2}>$ & $<P_{4}>$ \\
			 \hline\hline
			 18& 0.138 & -0.286& 0.649 & 0.463 & 0.474& -0.033\\
			 \hline
			 15& 0.073& -0.236& 0.629& 0.454 & 0.444& 0.025\\
			 \hline
			 10& 0.088 & -0.248 & 0.619 & 0.443 & 0.429& -0.081\\
			 \hline
			 5& 0.050& -0.218& 0.604& 0.437& 0.406 & 0.022\\
			 \hline
			 1& 0.480 & -0.494 & 0.519 & 0.342 & 0.279 & -0.075 \\
			 \hline
			\end{tabular}}
		\end{center}
		\hfill{}
		\caption{Temperature dependence of the order parameters of 5OO8 + 0.35\% wt CdSe-ZnS nanoparticles.} \label{top035}
	\end{table}

\begin{table}[ht]
\hfill{}
	\begin{center}
	{\small	\begin{tabular}{ccccccc}
			\hline
			$\Delta T\,[\mathrm{K}]$& a &b & $<\cos^{2}(\beta)>$&$<\cos^{4}(\beta)>$& $<P_{2}>$ & $<P_{4}>$ \\
			\hline\hline
			19.7 & 0.461 & -0.484 & 0.559 & 0.395 & 0.339 & 0.007\\
			\hline
			14.7& 0.669& -0.583& 0.511 & 0.332 & 0.267& -0.089\\
			\hline
			9.7 & 0.676 & -0.578 & 0.502 & 0.298 & 0.253 & -0.204\\
			\hline
			4.7& 0.688& -0.591& 0.433 & 0.259& 0.150 & -0.116\\
			\hline
			0.7& 0.963 & -0.700 & 0.432 & 0.259 & 0.148 & -0.199 \\
			\hline
		\end{tabular}}
	\end{center}
\hfill{}
\caption{Temperature dependence of the order parameters of 5OO8B + 0.4\% wt CdSe-ZnS nanoparticles.} \label{top05}
\end{table}

\begin{center}
\begin{table}[ht]
{\small		\begin{tabular}{ccccccc}
			\hline
			$\Delta T\,[\mathrm{K}]$& a & b & $<\cos^{2}(\beta)>$&$<\cos^{4}(\beta)>$& $<P_{2}>$ & $<P_{4}>$ \\
			\hline\hline
			19.3 & 0.624& -0.563& 0.524 & 0.302 & 0.286 & -0.269\\
			\hline
			14.3& 0.656& -0.577& 0.514 & 0.302 & 0.271& -0.231\\
			\hline
			9.3 & 0.960 & -0.699 & 0.490& 0.285 & 0.235 & -0.216\\
			\hline
			4.3& 1.318& -0.819& 0.415 & 0.161& 0.123 &-0.478\\
			\hline
			0.3& 0.406 & -0.455 & 0.353 & 0.207 & 0.030 & -0.043 \\
			\hline
		\end{tabular}}
\caption{Temperature dependence of the order parameters of 5OO8 + 1\% wt CdSe-ZnS nanoparticles.} \label{top10}
\end{table}
\end{center}




\end{document}